\documentclass[pre,twocolumn,superscriptaddress,showpacs,amsmath,amssymb]{revtex4}
\usepackage{graphicx}
\usepackage{dcolumn}
\usepackage{bm}

\begin{document}

\title{Ternary Social Networks: Dynamic Balance and Self-Organized Criticality}
\author{Qing-Kuan Meng}
\email{qkmeng@mail.bnu.edu.cn} \affiliation{Department of Physics,
Beijing Normal University, Beijing 100875, China}
\author{Wei Liu}
\email{wliuphys@gmail.com} \affiliation{School of Science, Xi'an
University of Science and Technology, Xi'an 710054, China}
\author{Jian-Yang Zhu}
\thanks{Author to whom correspondence should be addressed}
\email{zhujy@bnu.edu.cn} \affiliation{Department of Physics, Beijing
Normal University, Beijing 100875, China}

\date{\today}

\begin{abstract}
Antal \emph{et al.} [Phys. Rev. E \textbf{72}, 036121 (2005)] have
studied the balance dynamics on the social networks. In this paper,
based on the model proposed by Antal \emph{et al.}, we improve it
and generalize the binary social networks to the ternary social
networks. When the social networks get dynamically balanced, we
obtain the distributions of each relation and the time needed for
dynamic balance. Besides, we study the self-organized criticality on
the ternary social networks based on our model. For the ternary
social networks evolving to the sensitive state, any small
disturbance may result in an avalanche. The occurrence of the
avalanche satisfies the power-law form both spatially and
temporally. Numerical results verify our theoretical expectations.
\end{abstract}


\maketitle

\section{Introduction}

\label{sec1} Antal \emph{et al.} \cite{socialbalance} have studied
the balance dynamics on the social networks based on the notion of
\emph{social balance} \cite{balance1,balance2}. In the networks,
each node is connected to all the others, representing each person
knows all the others in the society. Each edge in the networks has
two values, +1 and -1. If the edge is +1, it means the two persons
are friendly with each other. If the edge is -1, it means the two
persons are hostile towards each other. At every step, they choose a
triangle randomly from the network. If the product of the three
edges is +1, the triangle is stable. Otherwise, if the product is
-1, the triangle is unstable. For the stable triangle, it satisfies
(i) the friend of my friend being my friend; (ii) the enemy of my
friend being my enemy; (iii) the friend of my enemy being my enemy;
and (iv) the enemy of my enemy being my friend. The unstable
triangles always try to be stable, but the final state of the
network depends on the edge flipping probability $p$, which is set
manually. If $p\ge \frac{1}{2}$, the network will reach the state of
"paradise", with all relations being friendly. Several studies
around the balance dynamics have been carried out, including the
studies of the university class of triad dynamics \cite{university}
and the satisfiability problem of computer science \cite{computer},
etc.

In the first half of this paper, at first based on Ref.
\cite{socialbalance}, we map the edge relations to the node
relations with some relaxation, so there are two opposing opinions
in the network. We suppose for each triad relation in the social
networks, it will change from stable to unstable or the reverse,
caused by the change of some persons' opinions in it. While for the
change, the person's opinion depending both on the other opinions in
the triad relation and the opinions all around him. Next,
considering in the social networks, there being two opposing
opinions is extreme, we introduce the neutral opinion. So the values
of each node are generalized to +1, 0 and -1, and the metastable
triad relations appear. We find the densities of each triad relation
associated with the Hamiltonian and the geometrical temperature
which is determined by the structure of the network without manual
setting. Based on our model, we obtain the distributions of each
triad relation and the time needed for dynamic balance both for the
binary and ternary cases.

In the latter half of this paper, we study the self-organized
criticality (SOC) \cite{soctheory1,soctheory2,soctheory3} on the
ternary social networks. The previous studies of the SOC on complex
networks are mainly based on the BTW model
\cite{soctheory1,socER,socSF1,socSF2,socSF3,socSW,RevMod}. In this
paper, we establish our model with some differences. As we know, in
the society, if one relation changes from one state into anther, it
may affect other relations associating with it. The effect may be
big or small, depending on many internal factors. So in this paper,
we simplify this phenomenon, with an eye to the triad balance
dynamics, and make two modifications compared to Sec. \ref{sec2}. We
find under specific conditions the SOC to be observable. Besides,
for the small-world network \cite{smallworld,addedge}, we find the
way of construction have an influence on the occurrence of the
avalanche. We analyze it theoretically. Numerical results verify our
theoretical expectations.

This paper is organized as follows. In Sec. \ref{sec2}, we map the
edge relations to the node relations and generalize the binary
networks to the ternary ones. For both the binary and the ternary
networks, we obtain the distributions of each triad relation and the
time needed for dynamic balance. In Sec. \ref{sec3}, we study the
SOC on the regular network and the small-world network respectively,
and find out the small-world effect on the occurrence of the
avalanche. We end this paper with conclusions in Sec. \ref{sec4}.

\section{Dynamic balance}
\label{sec2}
\subsection{The Node Model}
\label{nodemodel} 
\begin{figure}[tbp]
\includegraphics[clip,width=0.4\textwidth]{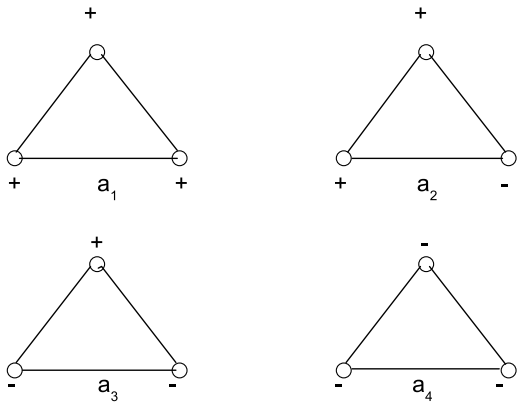}
\caption{\label{a1-a4} $a_1-a_4$ stand for each type of the triad
relations, and the corresponding densities in addition. $a_1$ and
$a_4$ are stable relations, while $a_2$ and $a_3$ are unstable
relations.}
\end{figure}

First we define the node relations in analogy with
Ref. \cite{socialbalance} with some relaxation. We propose each node
in the network has two values, +1 and -1, standing for each person's
two opposite standpoints in the society, or two opposing opinions.
For a triad relation, if all nodes have the same value, i.e., all
 persons have the same opinion, the relation is stable. If
there are different opinions in a triad relation, conflicts are
easily to occur, so the relation is unstable. The stable and
unstable triad relations are given in Fig. \ref{a1-a4}. In our
model, the stable and unstable triad relations have the
probabilities to change into the other, which is more realistic for
the social networks. As we know in the society, stable relations may
become unstable because of conflicts and the surroundings, and
unstable relations may become stable because of the same interest
and the surroundings. Considering each node has two values which
have the probabilities to change into the other, we study the social
networks in association with the spin systems. In the following
context, we establish our model in association with the Ising model
and the generalized Glauber dynamics \cite{glauber1,glauber2}.

When associated with the spin systems, each node in the network
standing for a spin. We propose the Hamiltonian of each triangle to be
\begin{equation}
H_{\Delta}=-\alpha(\sigma_{i}\sigma_{j}+\sigma_{j}\sigma_{k}+\sigma_{i}\sigma_{k}),\label{Hamilton}
\end{equation}
where $\sigma_i, \sigma_j$ and $\sigma_k$ stand for the three spins
of the triangle. In the binary case, $\sigma_i, \sigma_j$ and
$\sigma_k$ can take either of the two values, +1 and -1. Besides,
$\alpha>0$ is a real parameter, standing for the strength of
coupling. So the Hamiltonian is the form of the Ising model. The
evolving rules are as follows. At every step we choose a triangle
from the network randomly, and choose one of the three nodes from
the selected triangle at random, then let the selected node do spin
flipping. The spin flipping probability is defined as
\begin{equation}
p(\sigma_{i}\rightarrow \sigma'_{i})=\frac{e^{-\beta H'_{\Delta}}}{
Z},\label{probability}
\end{equation}
where
$H'_{\Delta}=-\alpha(\sigma'_{i}\sigma_{j}+\sigma_{j}\sigma_{k}+\sigma'_{i}\sigma_{k})$
stands for the final Hamiltonian after the spin flipping, with
$\sigma'_{i}$ the final value of the selected spin, and
$\sigma_{j}$, $\sigma_{k}$ the values of the other spins unchanged.
It is noted that the spin flipping probability depends on the final
value, independent of the initial one \cite{glauber2}. In addition,
$Z=\sum_{\sigma'_{i}=+1,-1}e^{-\beta H'_{\Delta}}$ is the
normalizing factor and $\beta$ is a geometrical temperature
determined by the structure of the network. Since $\beta$ is not a
real temperature, and in order to embody the majority principle, we
propose
\begin{equation}
\beta=1/|\sum_{j\neq i}\sigma_{j}/(N-1)-\sigma'_{i}|\simeq
 1/|M-\sigma'_{i}|,\label{temperature}
\end{equation}
where $N$ is the size of the network, and
$M=\sum_{i}\left<\sigma_{i}(t)\right>/N$ is the average value of all
spins. So the spin flipping probability depends both on the triad
relation and the surroundings \cite{glauber2,spinnet2}.

The numbers of positive and negative nodes in the network are as
follows,
\begin{eqnarray}
N^{+} &=&\frac{C_N^3\left( 3a_1+2a_2+a_3\right) }{C_N^2}=\frac
N3\left(
3a_1+2a_2+a_3\right) ,  \nonumber \\
N^{-} &=&\frac{C_N^3\left( a_2+2a_3+3a_4\right) }{C_N^2}=\frac
N3\left( a_2+2a_3+3a_4\right) .\label{number}
\end{eqnarray}
where $N$ is the number of total nodes and $a_i$ stands for the
triangle density, satisfying $\sum_{i=1}^{4}a_{i}=1$. The densities
of each triangle attached to a positive node are
\begin{equation}
n_{a_{i}}^{+}=\frac{\frac{N}{3}C_{i}a_{i}}{N^+}=\frac{C_{i}a_{i}}{3a_1+2a_2+a_3},
\end{equation}
where $i=1,2,3$, and $C_i$ is the number of positive nodes in $a_i$.
In a similar way, the densities of each triangle
 attached to a negative node are
\begin{equation}
n_{a_{j}}^{-}=\frac{\frac{N}{3}D_j
a_j}{N^-}=\frac{D_{j}a_{j}}{a_2+2a_3+3a_4},
\end{equation}
where $j=2,3,4$, with $D_j$ the number of negtive nodes in $a_j$.

For each node, the spin flipping probabilities from one value to the
other are given by Eq. (\ref{probability}), i.e.,
\begin{equation}
p\left( \sigma _i\rightarrow \sigma _i^{\prime }\right)
=\frac{e^{\frac \alpha {\left| M-\sigma _i^{\prime }\right| }\left(
\sigma _i^{\prime }\sigma _j+\sigma _j\sigma _k+\sigma _i^{\prime
}\sigma _k\right) }}{e^{-\frac \alpha {\left(
1+M\right) }\left( -\sigma _j+\sigma _j\sigma _k-\sigma _k\right) }+e^{\frac %
\alpha {\left( 1-M\right) }\left( \sigma _j+\sigma _j\sigma
_k+\sigma _k\right) }}.
\end{equation}
So the total flipping probabilities for each node from one value to
the other are as follows,
\begin{widetext}
\begin{eqnarray}
\overline{p}\left( 1\rightarrow -1\right)  &=&\sum_{i=1}^4\frac{C_ia_i}3%
p\left( 1\rightarrow -1\right)   \nonumber \\
&=&a_1\frac{e^{-\alpha /\left( 1+M\right) }}{e^{-\alpha /\left(
1+M\right) }+e^{3\alpha /\left( 1-M\right)
}}+\frac{2a_2}3\frac{e^{-\alpha /\left(
1+M\right) }}{e^{-\alpha /\left( 1+M\right) }+e^{-\alpha /\left( 1-M\right) }%
}+\frac{a_3}3\frac{e^{3\alpha /\left( 1+M\right) }}{e^{3\alpha
/\left( 1+M\right) }+e^{-\alpha /\left( 1-M\right) }},
\label{prob1}
\end{eqnarray}
\begin{eqnarray}
\overline{p}(-1 &\rightarrow
&1)=\sum_{i=1}^4\frac{D_ia_i}3p(-1\rightarrow 1)
\nonumber \\
&=&\frac{a_2}3\frac{e^{3\alpha /\left( 1-M\right) }}{e^{3\alpha
/\left( 1-M\right) }+e^{-\alpha /\left( 1+M\right)
}}+\frac{2a_3}3\frac{e^{-\alpha /\left( 1-M\right) }}{e^{-\alpha
/\left( 1-M\right) }+e^{-\alpha /\left( 1+M\right)
}}+a_4\frac{e^{-\alpha /\left( 1-M\right) }}{e^{-\alpha /\left(
1-M\right) }+e^{3\alpha /\left( 1+M\right) }},  \label{prob2}
\end{eqnarray}
\end{widetext}
where $a_{1}$ to $a_{4}$ stand for each triangle density, and $C_i$
and $D_i$ the number of positive and negative nodes in $a_i$
respectively. When the network gets dynamically balanced, the
densities of each triad relation are as follows,
\begin{equation}
a_1=\frac 18,~a_2=\frac 38,~a_3=\frac 38,~a_4=\frac 18.\label{den4}
\end{equation}
We note the network reaches an explicit state, independent of the parameter
$\alpha$. And in the dynamically balanced state $n^+=n^-$, where
$n^+$ and $n^-$ stand for the densities of positive and negative
nodes respectively.

The time needed for dynamic balance is
\begin{equation}
T_N\sim N^{C(\alpha)},
\end{equation}
where $N$ is the size of the network, and $C(\alpha)>0$, being a
function of $\alpha$. The numerical results are given in Fig.
\ref{TimeTwo}. More details are given in Appendix \ref{appen1}.
\begin{figure}[tbp]
\includegraphics[clip,width=0.4\textwidth]{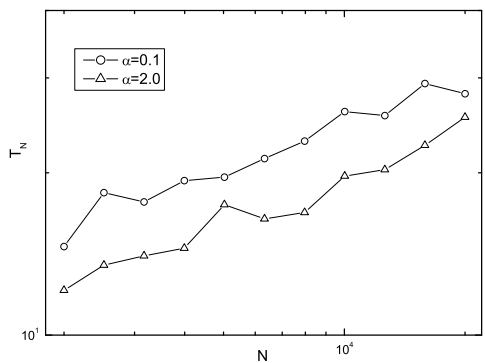}
\caption{\label{TimeTwo} A plot of $T_N$ with $N$ in the log-log
coordinate. $M(0)=-1$, $N$ is the size of the network, and $T_N$ is
the time needed for dynamic balance. The numerical results are
obtained by taking the average of $5\times 10^2$ simulations on the
networks with size from $2\times 10^3$ to $2\times 10^4$. }
\end{figure}

\subsection{The Generalized Node Model}
\begin{figure}[tbp]
\includegraphics[clip,width=0.4\textwidth]{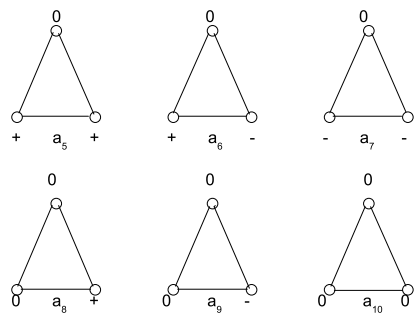}
\caption{\label{a5-a10} $a_5$ and $a_7$ represent the stable
relations, $a_6$ the unstable relation, while $a_8-a_{10}$ the
metastable relations.}
\end{figure}

There being only two opposing opinions in the social networks is
extreme. We propose the neutral opinion should exist, which means
the person holding this opinion is indifferent to other opinions. So
we propose each node should have three values, +1, 0 and -1,
standing for the three opinions. If there are the same opinion in
the triad relation except the neutral opinion, the relation is
stable. If there are different opinions in the triad relation except
the neutral opinion, the relation is unstable. Otherwise the
relation is metastable. The simplest way is by judging the
Hamiltonian (see Eq. (\ref{Hamilton})). If $H_{\Delta} <0$, it is
stable. If $ H_{\Delta} >0$, it is unstable. If $H_{\Delta}=0$, it
is metastable. There are four stable relations, three unstable
relations, and three metastable relations in the network, as shown
by Fig. \ref{a1-a4} and Fig. \ref{a5-a10}.

Compared with Eq. (\ref{number}), the numbers of positive, neutral
and negative nodes in the network are as follows,
\begin{eqnarray}
N^{+} &=&\frac N3\left( 3a_1+2a_2+a_3+2a_5+a_6+a_8\right), \nonumber \\
N^0 &=&\frac N3\left( a_5+a_6+a_7+2a_8+2a_9+3a_{10}\right), \nonumber \\
N^{-} &=&\frac N3\left(
a_2+2a_3+3a_4+a_6+2a_7+a_9\right),\label{threenum}
\end{eqnarray}
where $a_1$-$a_{10}$ stand for the densities of each triangle.
The densities of each triangle attached to a positive
node are as follows,
\begin{equation}
n_{a_{i}}^{+} =\frac{E_{i}a_{i}}{3a_1+2a_2+a_3+2a_5+a_6+a_8},
\end{equation}
where $i=1,2,3,5,6,8$, and $E_i$ is the number of positive nodes in
$a_i$.  In a similar way, the densities of each triangle attached to
a neutral node are as follows,
\begin{equation}
n_{a_{j}}^0 =\frac{F_{j}a_{j}}{a_5+a_6+a_7+2a_8+2a_9+3a_{10}} ,
\end{equation}
where $j=5,6,7,8,9,10$, and $F_j$ is the number of neutral nodes in
$a_j$. The densities of each triangle attached to a negative node
are as follows,
\begin{equation}
n_{a_{k}}^{-} =\frac{G_{k}a_{k}}{a_2+2a_3+3a_4+a_6+2a_7+a_9},
\end{equation}
where $k=2,3,4,6,7,9$, with $G_k$ being the the number of negtive
nodes in $a_k$.

We study the ternary network in association with the Potts spin
systems. Since the values of each spin are more than two, the spin
flipping mechanism changes into the spin transition mechanism. For
the spin transition mechanism, the spin takes one of the three
values, +1, 0 and -1, depending both on the triad relation and the
surroundings \cite{glauber2,spinnet2}. The Hamiltonian, the spin
transition probabilities and the geometrical temperature are
illustrated by Eqs. (\ref{Hamilton}), (\ref{probability}) and
(\ref{temperature}) respectively. The total probabilities from one
value to the others for each node are as follows,
\begin{widetext}
\begin{eqnarray}
\overline{p}\left( -1\rightarrow 0\right)  &=&\frac{a_2}3\frac{e^{\frac \alpha
{\left|
M\right| }}}{e^{\frac \alpha {\left| M\right| }}+e^{-\frac \alpha {1+M}}+e^{%
\frac{3\alpha }{1-M}}}+\frac{2a_3}3\frac{e^{-\frac \alpha {\left| M\right| }}%
}{e^{-\frac \alpha {\left| M\right| }}+e^{-\frac \alpha
{1+M}}+e^{-\frac \alpha {1-M}}}+a_4\frac{e^{\frac \alpha {\left|
M\right| }}}{e^{\frac \alpha {\left| M\right| }}+e^{\frac{3\alpha
}{1+M}}+e^{-\frac \alpha {1-M}}}
\nonumber \\
&&+\frac{a_6}3\frac 1{1+e^{-\frac \alpha {1+M}}+e^{\frac \alpha {1-M}}}+%
\frac{2a_7}3\frac 1{1+e^{\frac \alpha {1+M}}+e^{-\frac \alpha {1-M}}}+\frac{%
a_9}9,  \nonumber \\
\overline{p}\left( -1\rightarrow 1\right)  &=&\frac{a_2}3\frac{e^{\frac{3\alpha }{1-M}}%
}{e^{\frac \alpha {\left| M\right| }}+e^{-\frac \alpha {1+M}}+e^{\frac{%
3\alpha }{1-M}}}+\frac{2a_3}3\frac{e^{-\frac \alpha
{1-M}}}{e^{-\frac \alpha
{\left| M\right| }}+e^{-\frac \alpha {1+M}}+e^{-\frac \alpha {1-M}}}+a_4%
\frac{e^{-\frac \alpha {1-M}}}{e^{\frac \alpha {\left| M\right| }}+e^{\frac{%
3\alpha }{1+M}}+e^{-\frac \alpha {1-M}}}  \nonumber \\
&&+\frac{a_6}3\frac{e^{\frac \alpha {1-M}}}{1+e^{-\frac \alpha
{1+M}}+e^{\frac \alpha {1-M}}}+\frac{2a_7}3\frac{e^{-\frac \alpha {1-M}}}{%
1+e^{\frac \alpha {1+M}}+e^{-\frac \alpha {1-M}}}+\frac{a_9}9,  \nonumber \\
\overline{p}\left( 0\rightarrow -1\right)  &=&\frac{a_5}3\frac{e^{-\frac \alpha {1+M}}}{%
e^{-\frac \alpha {1+M}}+e^{\frac \alpha {\left| M\right|
}}+e^{\frac{3\alpha }{1-M}}}+\frac{a_6}3\frac{e^{-\frac \alpha
{1+M}}}{e^{-\frac \alpha
{1+M}}+e^{-\frac \alpha {\left| M\right| }}+e^{-\frac \alpha {1-M}}}+\frac{%
a_7}3\frac{e^{\frac{3\alpha }{1+M}}}{e^{\frac{3\alpha
}{1+M}}+e^{\frac
\alpha {\left| M\right| }}+e^{-\frac \alpha {1-M}}}  \nonumber \\
&&+\frac{2a_8}3\frac{e^{-\frac \alpha {1+M}}}{e^{-\frac \alpha
{1+M}}+1+e^{\frac \alpha {1-M}}}+\frac{2a_9}3\frac{e^{\frac \alpha {1+M}}}{%
e^{\frac \alpha {1+M}}+1+e^{-\frac \alpha {1-M}}}+\frac{a_{10}}3,
\nonumber
\\
\overline{p}\left( 0\rightarrow 1\right)  &=&\frac{a_5}3\frac{e^{\frac{3\alpha }{1-M}}}{%
e^{-\frac \alpha {1+M}}+e^{\frac \alpha {\left| M\right|
}}+e^{\frac{3\alpha }{1-M}}}+\frac{a_6}3\frac{e^{-\frac \alpha
{1-M}}}{e^{-\frac \alpha
{1+M}}+e^{-\frac \alpha {\left| M\right| }}+e^{-\frac \alpha {1-M}}}+\frac{%
a_7}3\frac{e^{-\frac \alpha {1-M}}}{e^{\frac{3\alpha
}{1+M}}+e^{\frac \alpha
{\left| M\right| }}+e^{-\frac \alpha {1-M}}}  \nonumber \\
&&+\frac{2a_8}3\frac{e^{\frac \alpha {1-M}}}{e^{-\frac \alpha
{1+M}}+1+e^{\frac \alpha {1-M}}}+\frac{2a_9}3\frac{e^{-\frac \alpha {1-M}}}{%
e^{\frac \alpha {1+M}}+1+e^{-\frac \alpha {1-M}}}+\frac{a_{10}}3,
\nonumber
\\
\overline{p}\left( 1\rightarrow -1\right)  &=&a_1\frac{e^{-\frac \alpha {1+M}}}{%
e^{-\frac \alpha {1+M}}+e^{\frac \alpha {\left| M\right|
}}+e^{\frac{3\alpha }{1-M}}}+\frac{2a_2}3\frac{e^{-\frac \alpha
{1+M}}}{e^{-\frac \alpha
{1+M}}+e^{-\frac \alpha {\left| M\right| }}+e^{-\frac \alpha {1-M}}}+\frac{%
a_3}3\frac{e^{\frac{3\alpha }{1+M}}}{e^{\frac{3\alpha
}{1+M}}+e^{\frac
\alpha {\left| M\right| }}+e^{-\frac \alpha {1-M}}}  \nonumber \\
&&+\frac{2a_5}3\frac{e^{-\frac \alpha {1+M}}}{e^{-\frac \alpha
{1+M}}+1+e^{\frac \alpha {1-M}}}+\frac{a_6}3\frac{e^{\frac \alpha {1+M}}}{%
e^{\frac \alpha {1+M}}+1+e^{-\frac \alpha {1-M}}}+\frac{a_8}9,  \nonumber \\
\overline{p}\left( 1\rightarrow 0\right)  &=&a_1\frac{e^{\frac \alpha {\left|
M\right|
}}}{e^{-\frac \alpha {1+M}}+e^{\frac \alpha {\left| M\right| }}+e^{\frac{%
3\alpha }{1-M}}}+\frac{2a_2}3\frac{e^{-\frac \alpha {\left| M\right| }}}{%
e^{-\frac \alpha {1+M}}+e^{-\frac \alpha {\left| M\right|
}}+e^{-\frac
\alpha {1-M}}}+\frac{a_3}3\frac{e^{\frac \alpha {\left| M\right| }}}{e^{%
\frac{3\alpha }{1+M}}+e^{\frac \alpha {\left| M\right| }}+e^{-\frac
\alpha
{1-M}}}  \nonumber \\
&&+\frac{2a_5}3\frac 1{e^{-\frac \alpha {1+M}}+1+e^{\frac \alpha {1-M}}}+%
\frac{a_6}3\frac 1{e^{\frac \alpha {1+M}}+1+e^{-\frac \alpha {1-M}}}+\frac{%
a_8}9.\label{TranProb}
\end{eqnarray}
\end{widetext}
But Eq. (\ref{TranProb}) is
correct only for $M\neq 0$. When $M\rightarrow 0$, the transition probabilities
are defined as (the explanations are given in Appendix \ref{appen1})
\begin{eqnarray}
\overline{p}\left( -1\rightarrow 0\right)  &=&0,~\overline{p}\left( 1\rightarrow 0\right)
=0,
\nonumber \\
\overline{p}\left( 0\rightarrow 1\right)  &=&\frac 12\left( \frac{a_5}3+\frac{a_6}3+%
\frac{a_7}3+\frac{2a_8}3+\frac{2a_9}3+a_{10}\right) ,  \nonumber \\
\overline{p}\left( 0\rightarrow -1\right)  &=&\frac 12\left( \frac{a_5}3+\frac{a_6}3+%
\frac{a_7}3+\frac{2a_8}3+\frac{2a_9}3+a_{10}\right) ,  \nonumber \\
\overline{p}\left( 1\rightarrow -1\right)  &=&\frac 12\left( a_1+\frac{2a_2}3+\frac{a_3}%
3+\frac{2a_5}3+\frac{a_6}3+\frac{a_8}3\right) ,  \nonumber \\
\overline{p}\left( -1\rightarrow 1\right)  &=&\frac 12\left( \frac{a_2}3+\frac{2a_3}%
3+a_4+\frac{a_6}3+\frac{2a_7}3+\frac{a_9}3\right) .\label{TranProb1} \nonumber \\
\end{eqnarray}

\begin{figure}[tbp]
\includegraphics[clip,width=0.4\textwidth]{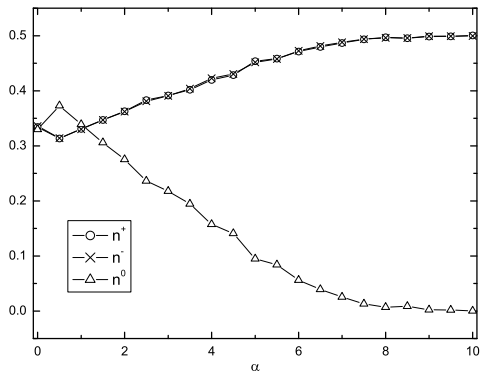}
\caption{\label{TimeThree} The numerical results are obtained by
taking the average of 20 simulations on the network with $10^4$
nodes. Initially $M(0)=-1/2, ~y(0)=1/2$, and the network takes a
long enough time to get dynamically balanced.}
\end{figure}

Since the transition probabilities are not continued at $M=0$, we
cannot give the analytical results of  $n^+, n^-, n^0$ with
$\alpha$, so we turn to the numerical method. The results are shown
by Fig. \ref{TimeThree}. We note when $\alpha\rightarrow 0$,
$n^+=n^-=n^0=1/3$. And when $\alpha \rightarrow +\infty, n^{0}=0$,
which means the neutral opinion disappears, and the ternary network
comes back to the binary network.

When the network gets dynamically balanced, the distributions of
each triangle are as follows,
\begin{equation}
a_{i}\simeq C_{lmn}x^{l+n}y^{m},
\end{equation}
where $n^+\simeq n^-=x$ and $n^0=y$. Besides, $i=1,2,3,...,10$, and
$l,m,n$ are the respective numbers of positive, neutral, negative
nodes in  $a_i$, along with $C_{lmn}$  the corresponding
combinatorial number. For example $C_{201}=3$.

For the time needed for dynamic balance, we expect it to satisfy
\begin{equation}
T_{N}\sim  N^{D(\alpha)},
\end{equation}
where $N$ is the size of the network and $D(\alpha)>0$, being a
function of $\alpha$.

The numerical results of the time needed for dynamic balance are
given in Fig. \ref{BalanceTimeThree}. More details are given in
Appendix \ref{appen2}.
\begin{figure}[tbp]
\includegraphics[clip,width=0.4\textwidth]{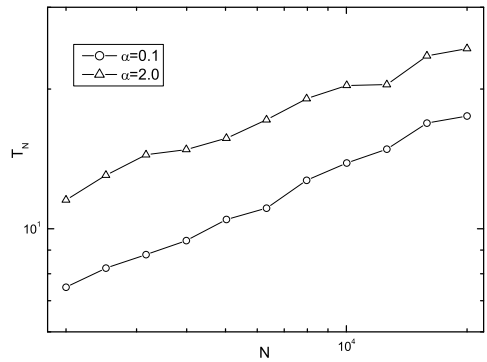}
\caption{\label{BalanceTimeThree} A plot of $T_N$ with $N$ in the
log-log coordinate. $M(0)=-1/2,~ y(0)=1/2$. $N$ is the size of the
network. $T_N$ is the time needed for dynamic balance. The numerical
results are obtained by taking the average of $5\times 10^2$
simulations on the networks with size from $2\times 10^3$ to
$2\times 10^4$. }
\end{figure}

\section{Self-Organized Criticality}
\label{sec3} As is well known, SOC
\cite{soctheory1,soctheory2,soctheory3} is studied on many complex
systems, including on the random graph \cite{socER}, the scale-free
network \cite{socSF1,socSF2,socSF3} and the small-world network
\cite{socSW}, etc, which are mainly based on the BTW model
\cite{soctheory1}. Besides, there are the studies of the SOC on
complex networks based on other models, which we refer the readers
to Ref. \cite{RevMod} for details. In this paper, we study the SOC
on the ternary social networks based on our model and the evolving
rules, along with two modifications compared to Sec. \ref{sec2}. The
first modification is the structure of the network. Because for a
completely connected network, any disturbance is globe, with no
propagation, so the occurrence of the avalanche will not follow the
power-law form either spatially or temporally. Besides, compared to
the completely connected network, the regular network or the
small-world network \cite{smallworld,addedge} is closer to the real
structure of the society. So we take the latter forms for study. The
second modification is the judgement of a triad relation. In order
to embody the majority principle, we take the judgement as the gauge
invariance or the gauge variance
\cite{gaugeinvary1,gaugeinvary2,spinnet1,spinnet2}, which is
explained as follows. Under this judgement, the triad relation
evolves from one state to the other by changing node's value.

If the value of a node being +1, it means the person shows the
friendly face in the triad relation, and always tires to keep the
relation stable. If the value being -1, it means the person shows
the hostile face, and always tries to destroy the stability of the
triad relation. While the value being 0, it means the person does
not care whether the triad relation is stable or not. So we propose
if the friendly faces are in the majority, the triad relation is in
one state, or the gauge invariance state. While the hostile faces
are in the majority, the triad relation is in the other state, or
the gauge variance state. The explicit expressions of the gauge
invariance and the gauge variance states
\cite{gaugeinvary1,gaugeinvary2,spinnet1,spinnet2} are as follows.

For each triad relation, we propose if
\begin{equation}
\sigma_{1}+\sigma_{2}+\sigma_{3}\geq 0~or
~\sigma_{1}=\sigma_{2}=\sigma_{3}=0,\label{gauge1}
\end{equation}
it satisfies the gauge invariance. If
\begin{equation}
\sigma_{1}+\sigma_{2}+\sigma_{3}<0~or~\sigma_{1}\neq \sigma_{2}\neq
\sigma_{3}=0,\label{gauge2}
\end{equation}
the triad relation does not satisfy the gauge invariance. Here
$\sigma_1, \sigma_2$ and $\sigma_3$ are the values of the three
nodes of the triangle. So $a_1,a_2,a_5,a_8,a_{10}$ are gauge
invariance states, while $a_3,a_4,a_6,a_7,a_9$ are gauge variance
ones, as shown by Fig. \ref{a1-a4} and Fig. \ref{a5-a10}.

We study the ternary network in association with the Potts spin
systems. Since each node has three values, the way of spin changing
is the spin transition \cite{glauber2}. Besides, we propose the spin
transition depends both on itself and the surroundings
\cite{glauber2,spinnet2}. Considering the SOC is determined by the
structure of the system, we set the geometric temperature as
$\beta=\frac{1}{T}>0$, where
$T=|\sum_{<i,j>}\sigma_{j}/n_{i}-\sigma'_{i}|$, with $\sigma'_i$ the
final value of $i$, and $\sum_{<i,j>}\sigma_{j}/n_{i}$ the average
value of all neighboring spins of $i$. Based on these
considerations, we give the spin transition probability of each node
as follows,
\begin{equation}\label{prob}
W(\sigma_{i}\rightarrow \sigma'_{i})=\frac{1}{Z}e^{-\beta
H(\sigma'_{i}, \Sigma_{<i,j>}\sigma_{j})}.
\end{equation}
In Eq. (\ref{prob}),
$H(\sigma'_{i},\sum_{<i,j>}\sigma_{j})=-\sigma'_{i}\sum_{<i,j>}\sigma_{j}$,
where $\sum_{<i,j>}\sigma_{j}$ is the sum of values of all
neighboring spins of $i$, and $Z=\sum_{\sigma'_{i}=-1,0,+1}e^{\beta
\sigma'_{i}\Sigma_{<i,j>}\sigma_{j}}$ is the normalizing factor. In
order to avoid the denominator of $\beta$ being zero, we propose if
the final value of the spin results in $T=0$, it happens with
probability zero, while the other values happen with
probability one. In this way, whatever the initial condition is, the
network will evolve to the sensitive state.

The evolving rules are as follows. We set all triad relations in the
network to satisfy gauge invariance initially. The simplest way is
to set all spins to be zero, which stands for all persons in the
society being neutral with each other initially. After a long enough
time of evolution, the network will reach the sensitive state. Then
any small disturbance may result in an avalanche.

For any small disturbance:

(1) Choose a triangle at random. If it is gauge variance, nothing
happens. If it is gauge invariance, we choose one of the three nodes
at random, and let the node do spin transition according to Eq.
(\ref{prob}). If after the spin transition, the triangle is still
gauge invariance, nothing happens. Otherwise, we store all the gauge
invariance triangles attached to the selected triangle, and goto
step (2).

(2) Choose all the stored triangles successively at random. Because
of the interaction, first we judge whether the triangle is gauge
invariance. If it is gauge variance, nothing happens. Otherwise, let
one of the three nodes do spin transition. If the final triangle is
gauge invariance, nothing happens, otherwise we find out all the
gauge invariance triangles attached to the selected triangle.

(3) Find out all the gauge invariance triangles by the combined
action of the stored triangles in step (2). If the number of gauge
invariance triangles to store is nonzero, goto step (2), otherwise
goto step (1).

For each avalanche, we define the size as the number of triangles
stored in step (2) changed from the gauge invariance state to the
other state, and plus the one changed in step (1). Besides, we
define the time of avalanche as follows. If the number of stored
triangles in step (2) changed from the gauge invariance state to the
other state is nonzero, the time is increased by one, and plus the
one changed in step (1).

\begin{figure}[tbp]
\includegraphics[clip,width=0.4\textwidth]{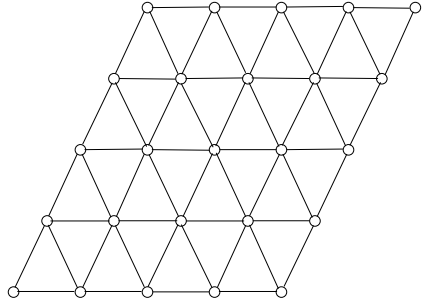}
\caption{\label{Hexangular} There are $9\times 10^4$ nodes in the
two-dimensional lattice with equal length and width. It is one of
the simplest two-dimensional regular networks satisfying the triad
relations. }
\end{figure}

\begin{figure}[tbp]
\includegraphics[clip,width=0.5\textwidth]{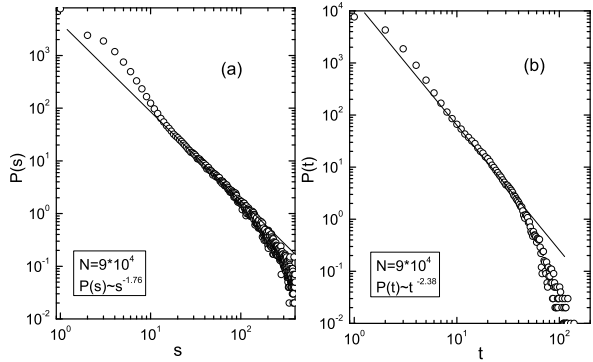}
\caption{\label{RegM-3} Plots of $P(s)$ versus $s$ and $P(t)$ versus
$t$ on the regular network. (a) $s$ is the size of avalanche. $P(s)$
is the distribution function, satisfying the power-law form, with
exponent -1.76. (b) $t$ is the time of avalanche. $P(t)$ is the
distribution function, satisfying the power-law form, with exponent
-2.38.}
\end{figure}

At first, we study the SOC on the regular network. The structure of
the regular network is given in Fig. \ref{Hexangular}. The size
distribution and the time distribution are given in Fig.
\ref{RegM-3} (a) and \ref{RegM-3} (b) respectively. In the following
content, we keep the number of nodes in both the regular network and
the small-world network constant as $9\times 10^4$. Besides, the
numerical results are obtained by taking the average of 100
simulations. For each simulation, we execute $9\times 10^6$ time of
disturbance.

In Fig. \ref{RegM-3} (a) and Fig. \ref{RegM-3} (b), the diagrams are
curved when $s$ and $t$ are small. Because we set all spins to be
zero initially. It takes a long time for the network to reach the
sensitive state. While the diagrams curved at the ends is caused by
the finite size effect.

\begin{figure}[tbp]
\includegraphics[clip,width=0.45\textwidth]{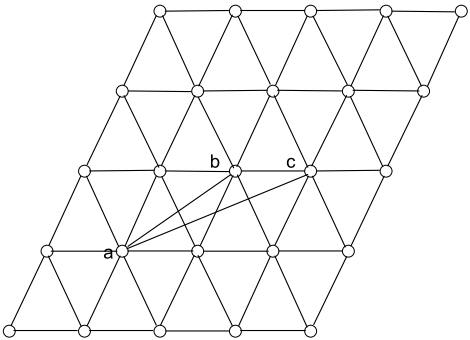}
\caption{\label{Smallworld} From a two-dimensional lattice with
equal length and width, for each node in the network, it sends out
an edge with probability $q$, and attaches to another node at
random, then attaches to one neighbor of the selected node at
random. Multi-connection and self-connection are avoided. Thus, by
adding edges, more triangles are created. The manner of adding
random edges is shown by nodes a,b,c.}
\end{figure}

Next, we study the SOC on the small-world network
\cite{smallworld,addedge}. The structure of the small-world network
is given in Fig. \ref{Smallworld}.

\begin{figure}[tbp]
\includegraphics[clip,width=0.5\textwidth]{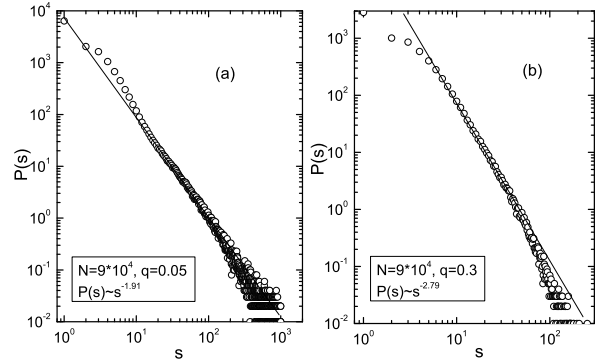}
\caption{\label{SWM-3} A plot of $P(s)$ versus $s$ on the
small-world network. $q$ is the edge adding probability. $s$ is the
size of the avalanche. $P(s)$ is the distribution function. (a)
$q=0.05$. The exponent of the power-law distribution is -1.91. (b)
$q=0.3$. The exponent of the power-law distribution is -2.79}
\end{figure}

When a triangle changes from gauge invariance to gauge variance, the
value of the selected node is decreased. Because of the small-world
effect, it may result in the triangles far away compared to the
regular network becoming of gauge variance. If the triangles far
away compared to the regular network become of gauge variance, it
will block the spread. So we expect the size of avalanche to
decrease, and the exponent of the power-law distribution to decrease
accordingly. Based on these analysis, we set the edge adding
probability $q$ of the small-world from $q=0.05$ to $q=0.3$, and
observe the exponent of the size distribution decrease from -1.91 to
-2.79, as given in Fig. \ref{SWM-3} (a) and Fig. \ref{SWM-3} (b)
respectively.

As is well known, for the small-world
network\cite{smallworld,addedge}, when the edge adding probability
$q$ is big enough, the structure of the network becomes random. We
note the diagrams are curved in Fig. \ref{SWM-3} (a) and Fig.
\ref{SWM-3} (b) when $s$ is small, but they take different forms. It
means that the random networks are easier for the spread in the
initial conditions.

\begin{figure}[tbp]
\includegraphics[clip,width=0.5\textwidth]{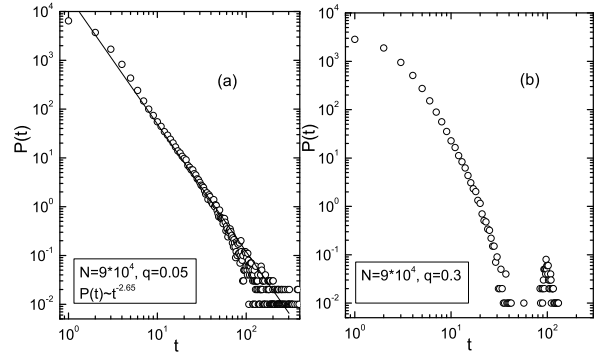}
\caption{\label{SWTMM-3} A plot of $P(t)$ versus $t$ on the
small-world network. $q$ is the edge adding probability. $t$ is the
time of avalanche. $P(t)$ is the distribution function. (a)
$q=0.05$. The exponent of the power-law distribution is -2.65. (b)
$q=0.3$.}
\end{figure}

Then there is the time distribution of the avalanche on the
small-world network, which is similar to the size distribution of
the avalanche, as given in Fig. \ref{SWTMM-3} (a) and Fig.
\ref{SWTMM-3} (b) for comparison. But we note, when $q$ is big
enough, with the structure of the network becoming random, the time
distribution no longer satisfies the power-law form. Besides, there
is condensation at the end of the diagram, which we want to study in
the later work.

\begin{figure}[tbp]
\includegraphics[clip,width=0.5\textwidth]{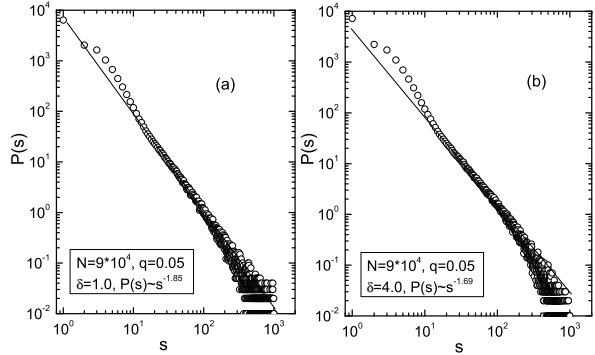}
\caption{\label{SWMD-3} A plot of $P(s)$ versus $s$ on the
small-world network considering the distance effect. $q$ is the edge
adding probability. $s$ is the size of avalanche. $P(s)$ is the
distribution function. (a) $q=0.05$, $\delta=1.0$, and the exponent
of the power-law distribution is -1.85. (b) $q=0.05$, $\delta=4.0$,
and the exponent of the power-law distribution is -1.69.}
\end{figure}

\begin{figure}[tbp]
\includegraphics[clip,width=0.5\textwidth]{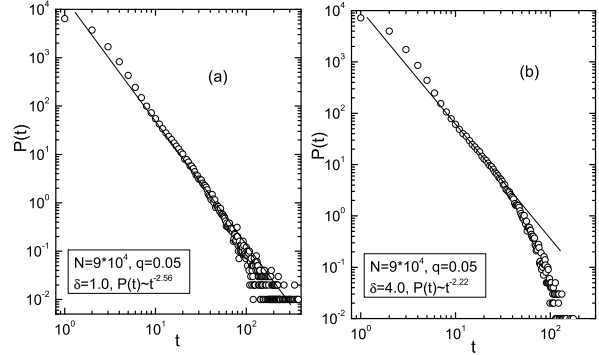}
\caption{\label{SWTMMD-3} A plot of $P(t)$ versus $t$ on the
small-world network considering the distance effect. $q$ is the edge
adding probability. $t$ is the time of avalanche. $P(t)$ is the
distribution function. (a) $q=0.05$, $\delta=1.0$, and the exponent
of the power-law distribution is -2.56. (b) $q=0.05$, $\delta=4.0$,
and the exponent of the power-law distribution is -2.22.}
\end{figure}

At last, for the construction of the small-world network, when
considering the factor of distance, the small-world effect on the
triangles far away compared to the regular network is weakened,
which may be easier for the spread. Besides, the number of local
triangles are increased. So we expect the size of avalanche to
increase, and the exponent of the size distribution to increase
accordingly. For the small-world network, when considering the
distance effect, it is constructed as follows. From a regular
network, as shown by Fig. \ref{Hexangular}, for each node in the
network, it sends out an edge with probability $q$ and attaches to
another node in the network. The probability of attachment between
nodes $i$ and $j$ is
\begin{equation}
p_{ij}=\frac{\ell^{-\delta}_{i,j}}{\sum_{m<n}\ell^{-\delta}_{m,n}}.
\end{equation}
Here $\delta>0$, and $\ell_{i,j}$ stands for the smallest distance
between $i$ and $j$ on the regular network. Then it attaches to one
neighbor of the selected node at random. Multi-connection and
self-connection are forbidden. In this manner, more triangles are
created. For the small-world network, we keep $q=0.05$ constant, and
set $\delta=1.0$ and $\delta=4.0$ for comparison. For the size
distribution of the avalanche, the exponent increases from -1.85 to
-1.69, as shown by Fig. \ref{SWMD-3} (a) and Fig. \ref{SWMD-3} (b).
Meanwhile, for the time distribution of the avalanche, the exponent
increases from -2.56 to -2.22, as shown by Fig. \ref{SWTMMD-3} (a)
and Fig. \ref{SWTMMD-3} (b).

In a word, the size distribution of the avalanche always satisfies
the power-law form, no matter on the regular network, the
small-world network or the random network. But for the time
distribution of the avalanche, it deviates from the power-law form
on the random network, which we want to do further study in the
later work.

\section{Conclusions}
\label{sec4} We have studied the social networks based on our model
and obtained meaningful results. First, in Sec. \ref{sec2} of this
paper, based on Ref. \cite{socialbalance}, we map the edge relations
to the node relations with some relaxation, along with introducing
the neutral relation. So the social networks change from the binary
networks into the ternary ones. We suppose the change of each triad
relation depending both on itself and the surroundings. So based on
our model, when the social networks get dynamically balanced, we
obtain the distributions of each triad relation and the time needed
for dynamic balance. Besides, we find under the extreme condition
$\alpha\rightarrow +\infty$, the ternary networks come back to the
binary ones. Second, in Sec. \ref{sec3} of this paper, we study the
SOC on the ternary social networks based on our model. Because we
suppose, if one triad relation changes, it will affect other triad
relations associating with it. The effect may be big or small,
depending on many internal factors. While compared with Sec.
\ref{sec2} of this paper, we make two modifications for our node
model as follows. One is the structure of the social networks, which
takes the form of regular or small-world network. The other is the
judgement of a triad relation, which takes the form of the gauge
invariance or the gauge variance. When the ternary social networks
evolving to the sensitive state, any small disturbance may result in
an avalanche. Since the number of triad relations changed in the
avalanche being no scale preferred, it should satisfy the power-law
distributions both spatially and temporally. Third, for the
small-world network, we find out the small-world effect on the
occurrence of the avalanche both theoretically and numerically.

\acknowledgments We thank H. Zhu for helpful discussions. The work
was supported by the National Natural Science Foundation of China
(No. 10875012).

\appendix

\section{Distributions of each triad relation and the time needed for dynamic balance}

\subsection{The node model}
\label{appen1} For the node model,  each node has two spin values,
+1 and -1. We define one step of time as executing $N$ spin flipping
successively at random \cite{socialbalance}. Because each spin
flipping is linked up with $(N-1)(N-2)/2$ variations of the
triangles, the variations of the densities of each triad relation with time
are as follows,
\begin{eqnarray}
\frac{da_1}{dt} &=&\overline{p}\left( -1\rightarrow 1\right)
n_{a_2}^{-}-\overline{p}\left(
1\rightarrow -1\right) n_{a_1}^{+},  \nonumber \\
\frac{da_2}{dt} &=&\overline{p}\left( 1\rightarrow -1\right)
n_{a_1}^{+}+\overline{p}\left(
-1\rightarrow 1\right) n_{a_3}^{-}  \nonumber \\
&&-\overline{p}\left( 1\rightarrow -1\right) n_{a_2}^{+}-\overline{p}\left( -1\rightarrow
1\right)
n_{a_2}^{-},  \nonumber \\
\frac{da_3}{dt} &=&\overline{p}\left( 1\rightarrow -1\right)
n_{a_2}^{+}+\overline{p}\left(
-1\rightarrow 1\right) n_{a_4}^{-}  \nonumber \\
&&-\overline{p}\left( 1\rightarrow -1\right) n_{a_3}^{+}-\overline{p}\left( -1\rightarrow
1\right)
n_{a_3}^{-},  \nonumber \\
\frac{da_4}{dt} &=&\overline{p}\left( 1\rightarrow -1\right)
n_{a_3}^{+}-\overline{p}\left( -1\rightarrow 1\right) n_{a_4}^{-}. \label{eqns}
\end{eqnarray}
When the network gets dynamically balanced, we have
\begin{equation}
\frac{da_1}{dt}=\frac{da_2}{dt}=\frac{da_3}{dt}=\frac{da_4}{dt}=0,\label{dadt0}
\end{equation}
and
\begin{equation}
\overline{p}(-1\rightarrow 1)=\overline{p}\left( 1\rightarrow -1\right)
\label{detailbalance}.
\end{equation}
From Eqs. (\ref{eqns}), (\ref{dadt0}) and (\ref{detailbalance}),
along with the normalization $\sum_{i=1}^{4}a_i=1$, we obtain
\begin{equation}
a_1=\frac 18,a_2=\frac 38,a_3=\frac 38,a_4=\frac 18.\label{den4}
\end{equation}
We note the distributions of each triad relation independent of the parameter $\alpha$, and
$n^+=n^-$ in the dynamically balanced state.

Defining at any time the density of positive nodes to be $\rho$, we
obtain
\begin{eqnarray}
a_{i}&=&C_{3}^{j}\rho^{j}(1-\rho)^{3-j}, \nonumber \\
 M&=&n^{+}-n^{-}=2\rho-1, \nonumber \\
\rho&=&(1+M)/2,~\rho(\infty )=\frac 12, \label{density}
\end{eqnarray}
where $j$ is the number of positive nodes in $a_i$ and $C_3^j$ is
the combinatorial number. We have the new equations as follows,
\begin{eqnarray}
\frac{dn^{+}}{dt} &=&\overline{p}(-1\rightarrow 1)-\overline{p}\left( 1\rightarrow
-1\right),
\nonumber \\
\frac{dn^{-}}{dt} &=&\overline{p}\left( 1\rightarrow -1\right) -\overline{p}(-1\rightarrow
1),
\nonumber \\
\frac{dM}{dt} &=&\frac{dn^{+}}{dt}-\frac{dn^{-}}{dt} \nonumber \\
&=&2[ \overline{p}(-1\rightarrow 1)-\overline{p}\left( 1\rightarrow
-1\right) ]. \label{mag1}
\end{eqnarray}
Substitute Eqs. (\ref{prob1}), (\ref{prob2}) and (\ref{density})
into Eq. (\ref{mag1}), we obtain
\begin{widetext}
\begin{eqnarray}
\frac{dM}{dt} &=&2\left[ \left( \frac{1+M}2\right) ^2\left( \frac{1-M}2%
\right) \frac{e^{3\alpha /\left( 1-M\right) }}{e^{3\alpha /\left(
1-M\right)
}+e^{-\alpha /\left( 1+M\right) }}+2\left( \frac{1+M}2\right) \left( \frac{%
1-M}2\right) ^2\frac{e^{-\alpha /\left( 1-M\right) }}{e^{-\alpha
/\left(
1-M\right) }+e^{-\alpha /\left( 1+M\right) }}\right.   \nonumber \\
&&+\left( \frac{1-M}2\right) ^3\frac{e^{-\alpha /\left( 1-M\right) }}{%
e^{-\alpha /\left( 1-M\right) }+e^{3\alpha /\left( 1+M\right)
}}-\left( \frac{1+M}2\right) ^3\frac{e^{-\alpha /\left( 1+M\right)
}}{e^{-\alpha
/\left( 1+M\right) }+e^{3\alpha /\left( 1-M\right) }}  \nonumber \\
&&\left. -2\left( \frac{1+M}2\right) ^2\left( \frac{1-M}2\right) \frac{%
e^{-\alpha /\left( 1+M\right) }}{e^{-\alpha /\left( 1+M\right)
}+e^{-\alpha
/\left( 1-M\right) }}-\left( \frac{1+M}2\right) \left( \frac{1-M}2\right) ^2%
\frac{e^{3\alpha /\left( 1+M\right) }}{e^{3\alpha /\left( 1+M\right)
}+e^{-\alpha /\left( 1-M\right) }}\right] \label{dM_dt1}
\end{eqnarray}
\end{widetext}
For Eq. (\ref{dM_dt1}), when $M(t) \sim 0$, after some calculations,
we obtain the approximate result,
\begin{equation}
M(t)\sim t^{-1/C(\alpha)},
\end{equation}
where $C(\alpha)>0$, being a function of $\alpha$.

For the dynamic balance of the network, we propose
\begin{equation}
\left| M\left( T_N\right) \right| \sim O\left( \frac 1N\right).
\end{equation}
So the time needed for dynamic balance is
\begin{equation}
T_N\sim N^{C(\alpha)}.
\end{equation}

In the simulations, both for the binary and the ternary cases,
if the final value $\sigma'_i$ results in the
denominator of $\beta$ (see Eq. (\ref{temperature})) being zero, we
propose it happens with probability zero, while the other values
happen with probability one (see Eq. (\ref{TranProb1}) for example).
In this manner, no matter what $M(0)$ (or $M(0)$ and $\rho(0)$)
is, the network will reach the dynamically balanced state.

\subsection{The generalized model}
\label{appen2} When each node has three spin values,
the variations of the densities of each triad relation with time are as
follows,
\begin{widetext}
\begin{eqnarray}
\frac{da_1}{dt} &=&\overline{p}\left( -1\rightarrow 1\right) n_{a2}^{-}+\overline{p}\left(
0\rightarrow 1\right) n_{a5}^0-\left[ \overline{p}\left( 1\rightarrow 0\right)
+\overline{p}\left(
1\rightarrow -1\right) \right] n_{a1}^{+},  \nonumber \\
\frac{da_2}{dt} &=&\overline{p}\left( 1\rightarrow -1\right) n_{a1}^{+}+\overline{p}\left(
-1\rightarrow 1\right) n_{a3}^{-}+\overline{p}\left( 0\rightarrow -1\right)
n_{a5}^0+\overline{p}\left( 0\rightarrow 1\right) n_{a6}^0 -\left[ \overline{p}\left(
1\rightarrow 0\right) +\overline{p}\left( 1\rightarrow -1\right) \right]
n_{a2}^{+}  \nonumber \\
&&-\left[ \overline{p}\left( -1\rightarrow 0\right) +\overline{p}\left(
-1\rightarrow 1\right) \right] n_{a2}^{-},  \nonumber \\
&&\vdots   \nonumber \\
\frac{da_9}{dt} &=&\overline{p}\left( 1\rightarrow 0\right) n_{a6}^{+}+\overline{p}\left(
-1\rightarrow 0\right) n_{a7}^{-}+\overline{p}\left( 1\rightarrow -1\right)
n_{a8}^{+}+\overline{p}\left( 0\rightarrow -1\right) n_{a10}^0 -\left[ \overline{p}\left(
0\rightarrow -1\right) +\overline{p}\left( 0\rightarrow 1\right) \right]
n_{a9}^0 \nonumber \\
&&-\left[ \overline{p}\left( -1\rightarrow 0\right) +\overline{p}\left(
-1\rightarrow 1\right) \right] n_{a9}^{-},  \nonumber \\
\frac{da_{10}}{dt} &=&\overline{p}\left( 1\rightarrow 0\right)
n_{a8}^{+}+\overline{p}\left( -1\rightarrow 0\right) n_{a9}^{-}-\left[ \overline{p}\left(
0\rightarrow -1\right) +\overline{p}\left( 0\rightarrow 1\right) \right]
n_{a10}^0  \label{eqn10s}.
\end{eqnarray}
\end{widetext}
When the network gets dynamically balanced, it satisfies
\begin{equation}
\frac{da_{i}}{dt}=0,\label{dadt10}
\end{equation}
where $i=1,2,3,...,10$,
along with the normalization
\begin{equation}
\sum_{i=1}^{10} a_{i}=1.\label{normal_10}
\end{equation}

Although Eqs. (\ref{eqn10s}) are very hard to solve, from appendix
\ref{appen1} and the symmetric spin transition mechanism, we expect $n^+\simeq n^-$
and $ M(\infty)\simeq 0$ in the dynamically balanced state. Defining
$n^+\simeq n^-=x$ and $n^0=y$  in the dynamically balanced state, the
densities of each type of the triangles are
\begin{equation}
a_{i}\simeq C_{lmn}x^{l+n}y^{m},
\end{equation}
where $l,m$ and $n$ are the numbers of positive, neutral and
negative nodes in $a_i$ respectively. $C_lmn$ is the corresponding
combinatorial number. For example $C_{201}=3$. Because near $M=0$,
there are fluctuations. The spin transition probabilities are
$(M\neq 0)$
\begin{widetext}
\begin{eqnarray}
\overline{p}\left( -1\rightarrow 0\right)  &\simeq &x^3+x^3+\frac{2x^2y}{1+e^\alpha
+e^{-\alpha }}+\frac{2x^2y}{1+e^\alpha +e^{-\alpha }}+\frac{xy^2}3,
\nonumber \\
\overline{p}\left( -1\rightarrow 1\right)  &\simeq &x^3+\frac{2x^2ye^\alpha
}{1+e^\alpha
+e^{-\alpha }}+\frac{2x^2ye^{-\alpha }}{1+e^\alpha +e^{-\alpha }}+\frac{xy^2}%
3,  \nonumber \\
\overline{p}\left( 0\rightarrow -1\right)  &\simeq &x^2y+\frac{2xy^2e^{-\alpha
}}{1+e^\alpha +e^{-\alpha }}+\frac{2xy^2e^\alpha }{1+e^\alpha
+e^{-\alpha }}+\frac{y^3}3,
\nonumber \\
\overline{p}\left( 0\rightarrow 1\right)  &\simeq &x^2y+\frac{2xy^2e^{-\alpha
}}{1+e^\alpha +e^{-\alpha }}+\frac{2xy^2e^\alpha }{1+e^\alpha
+e^{-\alpha }}+\frac{y^3}3,
\nonumber \\
\overline{p}\left( 1\rightarrow -1\right)  &\simeq &x^3+\frac{2x^2ye^{-\alpha
}}{1+e^\alpha +e^{-\alpha }}+\frac{2x^2ye^\alpha }{1+e^\alpha
+e^{-\alpha }}+\frac{xy^2}3,
\nonumber \\
\overline{p}\left( 1\rightarrow 0\right)  &\simeq &x^3+x^3+\frac{2x^2y}{1+e^\alpha
+e^{-\alpha }}+\frac{2x^2y}{1+e^\alpha +e^{-\alpha }} +\frac{xy^2}3, \nonumber \\
\end{eqnarray}
\end{widetext}
and ($M\rightarrow 0$)
\begin{eqnarray}
\overline{p}\left( -1\rightarrow 0\right)  &=&0,~\overline{p}\left( 1\rightarrow 0\right)
=0,
\nonumber \\
\overline{p}\left( 0\rightarrow 1\right)  &=&2x^2y+2xy^2+\frac{y^3}2,  \nonumber \\
\overline{p}\left( 0\rightarrow -1\right)  &=&2x^2y+2xy^2+\frac{y^3}2,  \nonumber \\
\overline{p}\left( 1\rightarrow -1\right)  &=&2x^3+2x^2y+\frac{xy^2}2,  \nonumber \\
\overline{p}\left( -1\rightarrow 1\right)  &=&2x^3+2x^2y+\frac{xy^2}2.
\end{eqnarray}

The densities of each triangle attached to a positive node are
\begin{equation}
n_{a_{i}}^{+}\simeq \frac{c_{1}x^2+2c_{2}x^2+2c_{3}xy+c_{4}y^2}{\left(
2x+y\right) ^2},
\end{equation}
where $i=1,2,3,5,6,8$, and $\sum_{j=1}^{4}c_{j}=1, c_{j}=0,1$. That
is, only one of the coefficient $c_{j}$ is nonzero. In a similar
way, the densities of each triangle attached to a neutral node are
\begin{equation}
n_{a_{j}}^0\simeq \frac{c_{1}x^2+2c_{2}x^2+2c_{3}xy+c_{4}y^2}{\left(
2x+y\right) ^2},
\end{equation}
where $j=5,6,7,8,9,10$. And the densities of each triangle attached
to a negative node are
\begin{equation}
n_{a_{k}}^{-}\simeq \frac{c_{1}x^2+2c_{2}x^2+2c_{3}xy+c_{4}y^2}{\left(
2x+y\right) ^2},
\end{equation}
where $k=2,3,4,6,7,9$.

Because the transition probabilities (see Eq. (\ref{TranProb}) and
Eq. (\ref{TranProb1})) are not continued at $M=0$, we cannot obtain
the analytical results for the distributions of $n^+, n^-$ and
$n^0$ with $\alpha$, so we turn to the numerical method, as shown by Fig. \ref{TimeThree}.
We note when $\alpha\rightarrow +\infty$, the neutral opinion disappears,
and the ternary network comes back to the binary network. That can be
explained as follows. In the triad relation, the more people try to keep it stable,
 i.e., to show the same opinion with the others, the less likely the
neutral opinion will exist.

\end{document}